\documentclass[superscriptaddress,10pt,preprintnumbers,showpacs,amsmath,amssymb,floatfix,twocolumn,prb]{revtex4}
\usepackage[utf8]{inputenc}
\usepackage{amsmath}
\usepackage{mathrsfs}
\usepackage{graphics}
\usepackage{xspace}
\usepackage[T1]{fontenc}
\usepackage{graphicx}% Include figure files
\usepackage{dcolumn}% Align table columns on decimal point
\usepackage{bm}% bold math
\usepackage{float}% add hypertext capabilities
\newcommand{\C}[1]{\hat c^{\dag}_{#1}}
\newcommand{\A}[1]{\hat c_{#1}}
\newcommand{\n}[1]{n_{#1}}
\newcommand{\ua}{\uparrow}
\newcommand{\da}{\downarrow}

\newcommand{\ra}{\rangle}
\newcommand{\la}{\langle}
\newcommand{\eref}[1]{Eq. (\ref{#1})}
\newcommand{\fref}[1]{Fig. \ref{#1}}

\begin{document}

\title{Low-energy effective theories of the two-thirds filled Hubbard model on the triangular necklace lattice}% Force line breaks with \\
\author{C. Janani}
\email{jananichander84@gmail.com}
\affiliation{Center for Organic Photonics and Electronics, School of Mathematics and Physics, University of Queensland, Brisbane, Queensland-4072, Australia}
\affiliation{Centre for Engineered Quantum Systems, School of Mathematics and Physics, University of Queensland, Brisbane, Queensland-4072, Australia.}
\author{J. Merino}
\affiliation{ Departamento de F\'{i}sica Te\'{o}rica de la Materia Condensada, Condensed Matter Physics Center (IFIMAC) and Instituto Nicol\'as Cabrera, Universidad Aut\'{o}noma de Madrid, Madrid 28049, Spain}
\author{Ian P. McCulloch }
\affiliation{Centre for Engineered Quantum Systems, School of Mathematics and Physics, University of Queensland, Brisbane, Queensland-4072, Australia.}
\author{B. J. Powell}\affiliation{Center for Organic Photonics and Electronics, School of Mathematics and Physics, University of Queensland, Brisbane, Queensland-4072, Australia}

\date{\today}% It is always \today, today,

\begin{abstract}

Motivated by   Mo$_3$S$_7$(dmit)$_3$, we investigate the Hubbard model on the  triangular necklace lattice at two-thirds filling. We show, using second order perturbation theory, that in the   molecular limit, the ground state and the low energy excitations of this model are identical to those of the  spin-one Heisenberg chain. The latter model is known to be in the symmetry protected topological Haldane phase.  Away from this limit we show, on the basis of symmetry arguments and density matrix renormalization group (DMRG) calculations, that the low-energy physics of the Hubbard model  on the triangular necklace lattice at two-thirds filling  is captured by the ferromagnetic  Hubbard-Kondo lattice  chain at half filling. This is consistent with and strengthens previous claims that both the half-filled  ferromagnetic  Kondo lattice  model and the two-thirds filled Hubbard model on the triangular necklace lattice are also in  the Haldane phase. A connection between Hund's rules and Nagaoka's theorem is also discussed.
\end{abstract}
\maketitle

\section{introduction\label{intro}} 

Geometric frustration has profound effects on the ground states and excitation spectra of low dimensional systems.\cite{Balents, Moessner} In two-dimensions it appears that very different physics can emerge on different frustrated lattices. For example, the Heisenberg model on the kagome lattice is a  spin-liquid, although it remains controversial whether it is gapped\cite{Yan} or not,\cite{Nakano,Iqbal} spin ice supports magnetic monopoles,\cite{ice}  and the anisotropic  triangular lattice shows a number of different phases with long-range order,  spin liquids and valence bond solid phases.\cite{RPP,Scriven0,Coldea} 

In one dimension,  the workhorse lattice for studying geometrical frustration is that  zig-zag   ladder. The spin $S=1/2$ Heisenberg model on the zig-zag lattice displays a number of exotic properties,\cite{White} for example, Majumdar and Ghosh proved that the ground state is a valence bond solid when $J_1=J_2/2$, where $J_1$ is the exchange interaction along a rung (or equivalently between nearest neighbours in a chain) and $J_2$ is the exchange interaction along a leg (next nearest neighbours in a chain). %Away from this exactly solvable point, bond order  and incommensurate spiral spin correlations are found.\cite{White} 
The spin-one Heisenberg model on the zig-zag lattice undergoes a transition from the  Haldane phase, for small $J_2$, to a phase with two intertwined strings each possessing string order for large $J_2$. In magnetic field,  both the  $S=1/2$ and $S=1$ zigzag Heisenberg models display vector chiral order.\cite{Ian} The Hubbard model on the zig-zag ladder is a Luttinger liquid for small values  of the rung hopping, but increasing the frustration drives the system to a quantum critical point where one section of the Fermi sea is destroyed.\cite{Daul,Hamacher}    

It is clear from the richness of frustrated models discussed above  that it is important to investigate additional frustrated models, particularly when they describe real materials.  One dimensional models are particularly valuable because of the range of high accuracy numerical and analytical techniques available to understand such systems. 

%In this paper,  we investigate the Hubbard  model on a  interesting lattice structure in one dimension which we dub the the triangular necklace lattice, which  consist of triangles connected to each other via one vertex (c.f. \fref{figure:sketch}).  This lattice with triangles is   an ideal test bed  to see the interplay of quantum frustration, quantum fluctuations and strong correlations. 

\begin{figure}
 \begin{center}
 \includegraphics[width = \columnwidth]{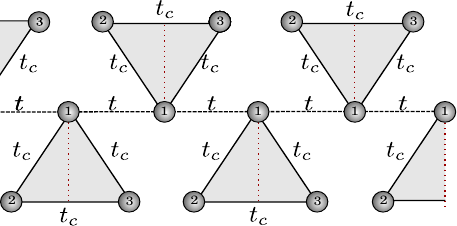}
\end{center}
\caption{(Color online.) The triangular necklace lattice. The Hubbard model on this lattice has two hopping terms. The three sites within each triangle are connected by a hopping integral $t_c$ (solid lines), and each triangle   is connected to its nearest neighbour  by a hopping integral $t$ between the 1-sites only (dashed line). The maroon dotted line marks the reflection equivalent to the local parity symmetry, i.e., relabeling sites 2 and 3 on any single molecule.}
\label{figure:sketch}
\end{figure}

It has previously been argued\cite{jacs,Janani} that that the triangular necklace  lattice (sketched in Fig. \ref{figure:sketch})  captures   the  underlying structure of Mo$_3$S$_7$(dmit)$_3$, where dmit is 1,3-dithiol-2-thione-4,5-dithiolate.  In this model, each triangular cluster represents   a molecule of Mo$_3$S$_7$(dmit)$_3$, with the lattice sites representing hybrid Mo-dmit orbitals. Experimentally, Mo$_3$S$_7$(dmit)$_3$ is an insulator that displays no magnetic order down to the lowest temperatures studied. Non-magnetic density functional calculations predict a metallic  state, and only find an insulator if long-range antiferromagnetism is counter factually assumed.\cite{jacs,JCDFT} Similarly the tight-binding model on the triangular necklace  lattice is metallic for parameters appropriate to Mo$_3$S$_7$(dmit)$_3$ (in particular two-thirds filling; see below for details). This suggests that electronic correlations may play an important role. Therefore we have previously argued\cite{Janani} that the simplest possible Hamiltonian that may describe Mo$_3$S$_7$(dmit)$_3$ is the Hubbard model the triangular necklace  lattice:
\begin{eqnarray}
\hat H&=&U\sum_{i\alpha} \hat n_{i\alpha\uparrow}\hat n_{i\alpha\downarrow}-t_c\sum_{i, \alpha \neq \beta, \sigma} \hat c^{\dag}_{i\alpha\sigma} \hat c_{i\beta\sigma}\nonumber \\ && 
 -t\sum_{i\sigma} \left(\hat c^{\dag}_{i1\sigma} \hat c_{(i+1)1\sigma} 
 +H. c.\right)  \label{eq:ham},
\end{eqnarray}
where $t_c$ is the is the intramolecular hopping integral, $t$ is the intermolecular hopping integral, $\hat c^{(\dag)}_{i\alpha\sigma}$ annihilates (creates) an electron with spin $\sigma$ on the $\alpha^\textrm{th}$ site of the  $i^\textrm{th}$ molecule, and $\hat n_{i\alpha\sigma}=\hat c^{\dag}_{i\alpha\sigma}\hat c_{i\alpha\sigma}$. 
Below, we study  this model with $t_c>0$ and   $n=4$ electrons per triangle (two-thirds filling), which are the relevant parameters for Mo$_3$S$_7$(dmit)$_3$.  This model has been previously\cite{Janani}  studied by density matrix renormalization group (DMRG) calculations. These calculations found that the model has an insulating ground state that supports a symmetry protected topologically spin liquid that is in the Haldane phase,\cite{Haldane1,Haldane2} i.e., adiabatically connected to the ground state of the spin-one Heisenberg chain.

In this paper, we show that the low-energy physics of the two-thirds filled Hubbard model is described, in appropriate limits, by two previously studied models. This provides simple physical pictures of the insulating phase and the behavior of the spin degrees of freedom.

Firstly, we show that in the molecular limit, $t\rightarrow0$, the  Hubbard model on triangular necklace  lattice  at two-thirds filling  the low lying  excitations of are spin excitations described by the antiferromagnetic spin-one Heisenberg chain  whose ground state is in the Haldane phase.  The Haldane phase is a   gapped,   symmetry protected topological phase with  non-local string order and  fractionalized edge states. \cite{ Kennedy1,Kennedy2,hidden,Haldane1, Haldane2}

% It is been understood that Haldane phase of the spin-1 Heisenberg model is the ground state of many fermionic Hamiltonians  where the charge fluctuations are suppressed. 

Secondly, we show that away from the molecular limit the ferromagnetic Kondo lattice model with an onsite repulsion between the itinerant electrons, which we will refer to as the ferromagnetic Hubbard-Kondo lattice model, describes the low-energy excitations of the two-thirds filled Hubbard model of the triangular necklace model. This provides a simple picture of the insulating state at two-thirds filling in the Hubbard model. 
It has previously been argued that both the ferromagnetic Kondo lattice model and the ferromagnetic Hubbard-Kondo lattice model have ground states in the Haldane phase.\cite{Tsunetsugu,Garcia,Yanagisawa} 

 This paper is organised as follows. After a brief discussion of numerical methods in Sec. \ref{DMRG}, in section \ref{tb}, we solve the model for the $U=t=0$, and  transform  the Hamiltonian into the eigenbasis of the $U=t=0$ solution. This is a crucial conceptual step in the derivation of effective Hamiltonians that follows. We also discuss the symmetries of the full Hamiltonian focusing on the `local parity' symmetry, that is a key ingredient in localizing the spins in the effective ferromagnetic Hubbard-Kondo lattice model.  In Sec. \ref{pt} we take recourse to second order perturbation theory and demonstrate that low lying spectrum of the Hubbard model on the triangular lattice corresponds to the two site spin-one Heisenberg model. Finally in Sec. \ref{kondo},  we show that   the model reduces to the   ferromagnetic Hubbard-Kondo lattice model at half filling. 

\section{Methods: Density matrix renormalization group \label{DMRG}}

Although the central results presented here are analytical it is useful to compare these with numerical calculations to explore the parameter ranges where various approximations are valid. To do so, we employ  DMRG\cite{White93} implemented using the matrix product states (MPS)  ansatz \cite{Schollwouck} and SU(2) symmetry \cite{Ian07} using the `MPS toolkit' and keeping up to $\chi=1000$ states. We consider both dimers (six sites) and extended chains (120 sites or 40 molecules). All calculations are performed at at two-thirds filling (four electrons per molecule).

For dimers  there are eight electrons on  six sites there are $C^{12}_8=495$ states, where $C^n_k$ is the binomial coefficient associated with choosing $k$ objects from a set of $n$. Therefore, for dimers, we are able to retain all of the physical states and the DMRG  exactly diagonalizes the Hamiltonian. 

%In section \sref{kondo}, we use DMRG for large lattice sizes (40 molecular sites)  to obtain the orbital filling in various orbitals. The ground states obtained for  this size, have states  up to m= 1000 states.  During the convergence of the  ground states, the states  kept varying  at different sweeps, from 50 to 1000 states. Quantities of interest are extrapolated in m . Error bars are estimated from the convergence behavior in m.

\section{Molecular limit ($t\rightarrow0$)  \label{tb}}

\subsection{Molecular orbital theory ($U=0$) \label{MOT}}

\begin{figure}
     \begin{center}
   \includegraphics[width=\columnwidth]{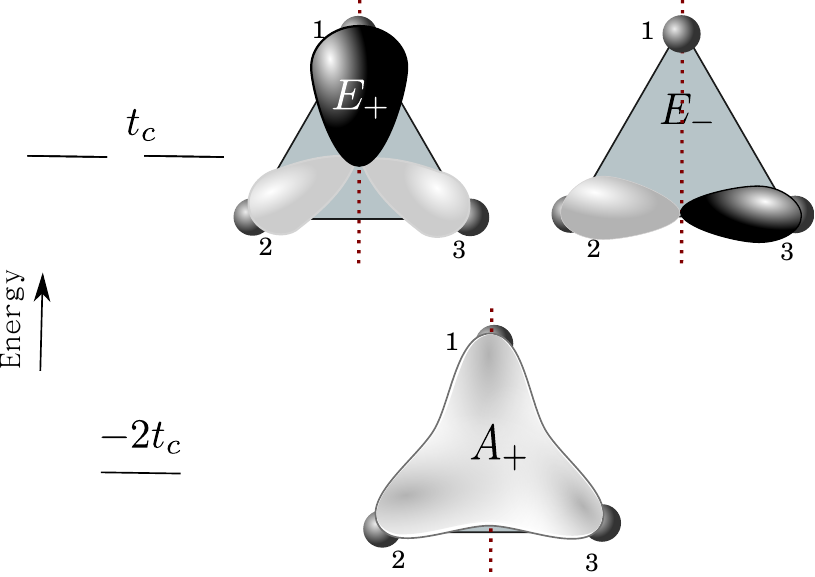}
\end{center}
 \caption{(Color online.) Sketches of the molecular orbitals for $t_c>0$.  The  different  colors on the orbitals imply different phases of the electron wavefunction. The orbitals   $A_{+}$ and $E_{+}$ have even parity under relabelling of sites 2 and 3, whereas $E_{-}$ has odd parity. Relabeling sites 2 and 3 is equivalent to reflecting (the $i^\textrm{th}$ molecule only) about the vertical maroon dotted line, which is also marked in Fig. \ref{figure:sketch}.}
 \label{orbitals}
    \end{figure}

The molecular limit,  $t\rightarrow0$, of Hamiltonian (\ref{eq:ham})  plays an important role in understanding the low-energy physics of this model. This is analogous  to the role of the atomic limit in  the Mott insulating phase of the half filled Hubbard model. However, the internal structure of the `molecule' (three site cluster) means that the molecular limit is not as straightforward as the atomic limit.   
When $U=t=0$,   Hamiltonian  (\ref{eq:ham}) reduces to the tight binding model on uncoupled triangular molecules:
\begin{equation}
\hat{\cal H}%= \sum_{i} \hat{\cal H}_{t_c}^{i}
= -t_c\sum_{i} \sum_{ \alpha \neq \beta, \sigma} \hat c^{\dag}_{i\alpha\sigma} \hat c_{i\beta\sigma}.\label{H_tc}
\end{equation}
It is straightforward to solve this Hamiltonian  for the $i^\textrm{th}$ molecule and  one  finds  three orbitals  with energies  $\varepsilon_{A_{+}}=-2t_c$, $\varepsilon_{E_{-}}=\varepsilon_{E_{+}}=t_c$. The corresponding wavefunctions, sketched in Fig. \ref{orbitals},  are given by
\begin{subequations}
\begin{eqnarray}
% |i,A_{+}\rangle&=& 
\hat c^\dagger_{iA_+\sigma}|0\rangle &\equiv &
\frac{1}{\sqrt{3}} (\hat c^\dagger_{i1\sigma}+\hat c^\dagger_{i2\sigma}+ \hat c^\dagger_{i3\sigma})|0\rangle, \label{A:orbital}\\
%|i,E_{-}\rangle&=& 
\hat c^\dagger_{iE_-\sigma}|0\rangle &\equiv & 
\frac{1}{\sqrt{2}} (\hat c^\dagger_{i2\sigma}-\hat c^\dagger_{i3\sigma})|0\rangle,\label{E-:orbital}
\end{eqnarray}
and
\begin{eqnarray}
%|i,E_{+}\ra=
\hat c^\dagger_{iE_+\sigma}|0\rangle &\equiv & 
\frac{1}{\sqrt{6}} (2\hat c^\dagger_{i1\sigma}-\hat c^\dagger_{i2\sigma}-\hat c^\dagger_{i3\sigma})|0\rangle,
\label{E+:orbital}
\end{eqnarray}
 \end{subequations}
where $|0\rangle$ is the vacuum state. We will refer to $\{\hat c^\dagger_{iA_+\sigma}|0\rangle, \hat c^\dagger_{iE_+\sigma}|0\rangle, \hat c^\dagger_{iE_-\sigma}|0\rangle\}$  as the molecular orbital basis and $\{\hat c^\dagger_{i1\sigma}|0\rangle, \hat c^\dagger_{2\sigma}|0\rangle, \hat c^\dagger_{i3\sigma}|0\rangle\}$  as the atomic orbital basis. The molecular orbitals  are labelled based on the C$_3$ symmetry of an isolated molecule and the parity of the wavefunction under exchange of sites 2 and 3 on any individual molecule which remains a symmetry of the Hamiltonian even for $t\ne0$, alternatively this transformation can be thought of as reflection through the maroon dotted lines in Figs. \ref{figure:sketch} and \ref{orbitals}. Henceforth, we will refer to the latter symmetry as local parity.

For $n=4$ and $t_c>0$,  which are the relevant  parameters of Mo$_3$S$_7$(dmit)$_3$, the ground state of the non-interacting ($U=0$) molecular limit corresponds to a  filled $A_{+}$ orbital and two electrons  shared between the two $E$ orbitals (\emph{i.e.}, the ground state is $C^4_2=6$-fold degenerate).  
For $t_c<0$ and $n=4$, the ground state corresponds to  filled $E$ orbitals and an empty $A_{+}$ orbital.

\subsection{Interactions in the molecular orbital basis \label{MO}}
We will see below that it  is  helpful to transform the Hamiltonian into the basis of the molecular orbitals. %The operators in \eref{eq:ham}  transform as
% \begin{eqnarray}
% \hat c_{iA_+\sigma}&=&(\hat c_{i1\sigma}+\hat c_{i2\sigma}+ \hat c_{i3\sigma})/\sqrt{3}, \nonumber\\
% \hat c_{iE_-\sigma}&=&(\hat c_{i2\sigma}-\hat c_{i3\sigma})/\sqrt{2},\nonumber\\
%\hat c_{iE_+\sigma}&=&(2\hat c_{i1\sigma}-\hat c_{i2\sigma}-\hat c_{i3\sigma})/ \sqrt{6},
%\end{eqnarray}
After this transformation the Hamiltonian \eref{eq:ham} can be rewritten in the form
\begin{equation}
\hat{\mathcal{H}}=\hat{\mathcal{H}}_\textrm{m}+\hat{\mathcal{H}}_t,
\end{equation}
where
\begin{eqnarray}
\hat{\mathcal{H}}_t
=\sum_{imn\sigma}\C{im\sigma}T_{mn}\A{(i+1)n\sigma}
+H.c.,
\label{perturbation}
\end{eqnarray}
$m, n\in\{ A_{+}, E_{+}, E_{-}\}$, $T_{mn}$ is the intermolecular  hopping matrix with matrix elements  $T_{A_+A_+}=-t/3$, $T_{A_+E_+}=T_{E_+A_+}=-\sqrt{2}t/{3}$, and  $T_{E_+E_+}=-{2}t/{3}$.  It can be seen from  \eref{E-:orbital} that $E_{-}$ orbitals has no weight on site $|1\ra$, so  $T_{E_{-} m}=0$ for any $m$. However, more pertinently, this is a direct consequence of the local parity symmetry (see Sec. \ref{parity}).
\begin{equation}
\hat{\cal H}_\textrm{m}=\hat{\cal H}_1
+\hat{\cal H}_2
+\hat{\cal H}_3,
\label{Ham:orb}
\end{equation}
 where, $\hat{\cal H}_n$ describes the interactions involving $n$ orbitals on a single molecule.  
\begin{equation}
\hat{\cal H}_1=  \sum_{im\sigma}\varepsilon_m \C{im\sigma}\A{im\sigma}+ U_m n_{im\ua} n_{im \da}, \label{H1}
\end{equation}
  where $U_{A_{+}}= U/3$ and $U_{E_{-}}= U_{E_{+}}=U/2$ . 
%Interactions involving  pairs of  orbitals are given by
\begin{widetext}
\begin{eqnarray}
 \hat{\cal H}_2=\sum_{m\ne n}\left[J_{mn}\sum_{i}\hat{\bf S}_{im}\cdot\hat{\bf S}_{in}+V_{mn }\sum_{i\sigma\sigma'}n_{im\sigma}n_{in\sigma'}%\nonumber\\&&
 +P_{mn}\sum_{i}\C{i m \ua }\C{i m\da}\A{i n\ua}\A{i n\da}%\nonumber\\&&
 +X_{mn}\sum_{i\sigma}\left(n_{im\sigma}\C{im\bar{\sigma} }\A{in\bar{\sigma}}+H.c.\right)\label{Hmn}\right], \label{H2}
\end{eqnarray}
where  $\hat{\bf S}_{im}=\C{im\alpha}{\bm{\sigma}}_{\alpha\beta}\A{im\beta}$, ${\bm{\sigma}}$ is the vector of Pauli matrices, $J_{mn}$ is the ferromagnetic interorbital exchange interaction, $V_{mn}$ is the interorbital Coulomb interaction, $P_{nm}$ is a two electron interorbital hopping, and $X_{mn}$ is a correlated interorbital hopping. The Hermiticity of the Hamiltonian requires that $J_{mn}= J_{nm}$, $V_{mn}= V_{nm}$,   and  $P_{mn}= P_{nm}$, the C$_3$ symmetry of the isolated molecule requires that $J_{mE_+}=J_{mE_-}$, $V_{mE_+}=V_{mE_-}$, $P_{mE_+}=P_{mE_-}$; and the local parity symmetry requires that  $X_{mE_-}=X_{E_{-}m}=0$. Explicitly transforming from the atomic orbital basis to the molecular orbital basis reveals that the remaining undefined parameters are 
 $J_{A_{+} E_{+}}= -U/3$,  $J_{E_{+} E_{-}}=-U/6$,  $V_{A_{+} E_{+}}= U/12$,  $V_{E_{+} E_{-}}=U/24$,   $P_{A_{+} E_{+}}= -U/3$, $P_{E_{+} E_{-}}=-U/6$, $X_{A_{+} E_{+}}=0$ and  $X_{E_{+} A_{+}}=U/3\sqrt{2}$.  The problem of determining such parameters for first principles in molecular solids has recently been discussed extensively.\cite{Nakamura,Laura,Scriven1,Scriven2} 
 
%Similarly, the interactions involving all three orbitals $\hat{H}_{\mathbf{m}\mathbf{n}\mathbf{l}}$, is given by

\begin{eqnarray}
\hat{\cal H}_3&=&
\frac{U}{3 \sqrt{2}}\sum_{i\sigma}\left(\C{i A_{+}\sigma}\C{i E_{+}\overline\sigma}\A{i E_{-}\sigma}\A{i E_{-}\overline\sigma}+H.c\right)%\nonumber\\&&
+\frac{U}{3 \sqrt{2}}\sum_{i\sigma}\left(\C{i E_{-}\sigma}\A{i E_{-}\overline\sigma} \C{i E_{+}\overline\sigma}\A{i A_{+}\sigma} + H.c.\right)\nonumber\\&&%correlated hopping
  -\frac{U}{3\sqrt{2}}\sum_{i\sigma}\left(\n{i E_{-}\sigma}\C{i E_{+}\overline\sigma}\A{i A_{+}\overline\sigma} +H.c.\right) \label{H3}
\end{eqnarray}

Note that $\hat{\cal H}_\textrm{m}$  is the Hubbard model on a triangle, which  can be solved exactly.  The  ground state of this model  at two-thirds filling for $t_c>0 $ is a triplet with energy $-2t_c+U$.   The (degenerate) ground state wavefunctions for the $i^\textrm{th}$ monomer  are therefore 
\begin{subequations}
\begin{eqnarray}
|\phi_{i}^\Uparrow\ra &=& |A_+^{\ua \da},E_-^{\ua},E_+^{\ua}\ra = \C{iA_{+}\uparrow}\C{iA_{+}\downarrow}\C{iE_{-}\uparrow}\C{iE_{+}\downarrow}|0\ra \\
|\phi_{{i}}^\Downarrow\ra  &=&|A_+^{\ua \da},E_-^{\da},E_+^{\da}\ra = \C{iA_{+}\uparrow}\C{iA_{+}\downarrow}\C{iE_{-}\downarrow}\C{iE_{+}\uparrow}|0\ra \\
|\phi_{{i}}^0\ra &=&\frac{1}{\sqrt{2}}\left( 
|A_+^{\ua \da},E_-^{\ua},E_+^{\da}\ra + |A_+^{\ua \da},E_-^{\da},E_+^{\ua}\ra \right)%\nonumber\\
                      =\frac{1}{\sqrt{2}}\left( \C{iA_{+}\uparrow}\C{iA_{+}\downarrow}\C{iE_{-}\uparrow}\C{iE_{+}\uparrow}+
 \C{iA_{+\uparrow}}\C{iA_{+}\downarrow}\C{iE_{-}\downarrow}\C{iE_{+}\downarrow}\right)|0\ra,
\end{eqnarray}\label{psi:triangle}
\end{subequations}
\end{widetext}
 where the superscripts on the terms between the two equality signs label the spin(s) of the electron(s) in that orbital. For two-thirds filling, {\it i.e.},  four electrons in three orbitals, one of orbitals  will be doubly occupied.  It can be seen from the ground states  that  for $t_c, U>0$ and $t=0$, the $A_{+}$ orbital continues to be doubly occupied, as in the case of $U=0$.  This is not unexpected as $U_{A_{+}}=U/3<U_{E_{+}}=U_{E_{-}}=U/2$.  The remaining two electrons occupy $E_{+}$ and $E_{-}$ orbitals with  one electron each respectively as the intraorbital Coulomb interaction for the $E$ orbitals is greater than the competing interorbital interaction $V_{E_{+}E_{-}}=U/12$. The interaction $J_{E_{+} E_{-}}$ lowers the energy of the triplet relative to that of the singlet. Thus in the molecular limit the Hubbard model on the triangular necklace lattice consists of    spin triplet  molecules. 
 
\subsection{Nagaoka'a theorem and Hund's rules} 

The above result, that the grounds state of an isolated triangle is a triplet, can be understood in terms of two physical effects that are usually regarded as entirely separate pieces of physics: Nagaoka's theorem and Hund's rules. The most general formulation of Nagaoka's theorem \cite{Tasaki} states that for $U=\infty$ the ground state of the Hubbard model with $N-1$ electron on $N$ lattice sites has the maximum possible spin, $S=(N-1)/2$, if all of the intersite hopping integrals are negative.\cite{foot-sign} To make connection with this result we must return to the atomic orbital basis and make a particle-hole transformation, $\hat h_{i\alpha\sigma}=\hat c^\dagger_{i\alpha\sigma}$. The Hamiltonian for $t=0$ and $t_c>0$ is then
\begin{eqnarray}
\hat{\cal H}&=&-(-t_c)\sum_{i,\alpha\ne\beta,\sigma}\hat h^\dagger_{i\alpha\sigma}\hat h_{i\beta\sigma}\notag\\&&+U\sum_{i\alpha}\hat h^\dagger_{i\alpha\uparrow}\hat h_{i\alpha\uparrow}\hat h^\dagger_{i\alpha\downarrow}\hat h_{i\alpha\downarrow}+U.
\end{eqnarray}
We remove the trivial term by shifting the zero of energy: $\hat{\cal H}\rightarrow\hat{\cal H}-U.$ We then regain a Hubbard model with 2 fermions on 3 sites and all hopping integrals are negative. Thus Nagaoka's theorem implies an $S=1$ ground state when $U\rightarrow\infty$, as we found explicitly above. Of course, on this finite lattice the triplet groundstate is found even away from $U=\infty$, which is not guaranteed by Nagaoka's theorem. Furthermore, Nagaoka's theorem states that the ground state wavefunction contains only positive coefficients when written in the natural real-space many-body basis.\cite{Tasaki} Returning to the molecular orbital basis and working with electrons rather than holes, this statement is equivalent to the prediction that the $A_+$ orbital will be doubly occupied, as indeed we found explicitly above. In the molecular orbital basis it is clear that $J_{E_{+}E_{-}}$ plays a key role in stabilizing the Nagaoka state.

If we accept  that the $A_+$ orbital will be doubly occupied, then, the triplet ground state is what one would expect from the molecular Hund's rules applied to the $E$ orbital subspace. Note however, that we do not include an explicit Hund's rule coupling, rather one simply finds $J_{E_{+}E_{-}}<0$ on transforming into the molecular orbital basis. Indeed, the connection to Hund's rules goes beyond the maximization of $S$ within the $E$ manifold. One can rewrite the Hamiltonian in terms of   `molecular Kanamori parameters' $\tilde U, \tilde U', \tilde J$ and $\tilde J'$ (cf. Eq. (35) of Ref. \onlinecite{Hotta}). In which case the electron-electron interactions within the $E$ manifold of the $i^\textrm{th}$ molecule are given by
\begin{eqnarray}
{\cal H}^{E}_{i}&=&\tilde U\sum_{\nu} \hat n_{i\nu\uparrow}\hat n_{i\nu\downarrow}
+\tilde U'\sum_{\sigma\sigma'}\hat n_{iE_+\sigma}\hat n_{iE_-\sigma'}\notag\\
&&+\tilde J\sum_{\sigma\sigma'}\hat c^\dagger_{iE_+\sigma}\hat c^\dagger_{iE_-\sigma'}\hat c_{iE_+\sigma'}\hat c_{iE_-\sigma}\notag\\
&&+\tilde J'\left(\hat c^\dagger_{iE_+\uparrow}\hat c^\dagger_{iE_+\downarrow}\hat c_{iE_-\downarrow}\hat c_{iE_-\uparrow}+H.c.\right).
\end{eqnarray}
On writing Eq. (\ref{H2}) in this form one finds that $\tilde U=U/2$ and $\tilde U'=\tilde J=\tilde J'=U/6$. This satisfies two important constraints, which the Kanamori parameters are required to respect: (i) $\tilde J=\tilde J'$ and (ii) $\tilde U=\tilde U'+\tilde J+\tilde J'$. In terms of the parameters used elsewhere in this paper these constraints correspond to (i) $J_{E_+E_-}=P_{E_+E_-}$ and (ii) $U_{E_{+}}=U_{E_{-}}=2V_{E_{+}E_{-}}-\frac32J_{E_{+}E_{-}}-P_{E_{+}E_{-}}$.

Furthermore, it is the interactions described by the Kanamori parameters that are responsible for Hund's rules in the $e_g$ manifold of a $d$-electron systems.\cite{Hotta} Therefore, in transforming to the molecular orbital basis, we have explicitly derived Hund's rules for the $E$ manifold of the molecule.

Therefore, it is clear that, in this system at least, Hund's rules and Nagaoka's theorem result from the same underlying physics. It is natural to speculate that this connection may be more general.

\subsection{Local parity \label{parity}}

It is important to note that, even if we restrict the relabeling of sites 2 and 3 to a single molecule, the local parity (defined in Sec. \ref{MOT}) is a symmetry of the Hubbard model  for all $t, U$. Thus, all eigenstates, and in particular the ground state, have a definite local parity on every site individually. For example, in Sec. \ref{MOT} we saw that for $t=U=0$ and $t_c>0$ there are six degenerate ground states on each molecule, which we can label 
\begin{subequations}
\begin{eqnarray}
|\alpha_+\ra&=&|A_+^{\uparrow\downarrow}, E_-^{\uparrow\downarrow}, E_+^{0}\ra,\\ 
|\beta_-\ra&=&|A_+^{\uparrow\downarrow}, E_-^{\uparrow}, E_+^{\uparrow}\ra,\\ 
|\gamma_-\ra&=&|A_+^{\uparrow\downarrow}, E_-^{\uparrow}, E_+^{\downarrow}\ra,\\ 
|\delta_-\ra&=&|A_+^{\uparrow\downarrow}, E_-^{\downarrow}, E_+^{\uparrow}\ra,\\ 
|\epsilon_-\ra&=&|A_+^{\uparrow\downarrow}, E_-^{\downarrow}, E_+^{\downarrow}\ra,
\end{eqnarray} 
and
\begin{eqnarray}
|\zeta_+\ra&=&|A_+^{\uparrow\downarrow}, E_-^{0}, E_+^{\uparrow\downarrow}\ra.
\end{eqnarray}
\end{subequations}
  It is clear that $|\alpha_+\ra$ and $|\zeta_+\ra$ have even parity and $|\beta_-\ra$, $|\gamma_-\ra$, $|\delta_-\ra$, and $|\epsilon_-\ra$ have odd parity as the only molecular orbital with odd local parity with respect to the $i^\textrm{th}$ molecule is $\hat c_{iE_-\sigma}^\dagger|0\ra$. This means that  arbitrary perturbations that respect the local parity symmetry may mix $|\alpha_+\ra$ with $|\zeta_+\ra$ or any of the set $|\beta_-\ra$, $|\gamma_-\ra$, $|\delta_-\ra$, and $|\epsilon_-\ra$, but perturbations that respect the local parity symmetry will not mix even parity states with odd parity states. 

In particular, as non-zero $t$ and non-zero $U$ do not break local parity symmetry this means that even away from the $t=U=0$ limit the local eigenstates have a definite local parity with respect to each molecule individually. This is turn, implies that in any state the occupation number of the $E_-$ orbital is conserved modulo two (individually) on every molecule (as changes by $\pm1$ result in a change in the local parity). In particular when $n_{iE_-}\equiv\sum_\sigma n_{iE_-\sigma}=1$ the electron is localized on the $i^\textrm{th}$ molecule as long as there is no phase transition.

It is clear from  Eqs. (\ref{H1}-\ref{H3}) and the fact that $T_{mE_-}=0$ for all $m$ that the explicit form of $\hat{\cal H}$ given in Sec. \ref{MO} conserves $\hat n_{iE-}$ modulo two for all $i$.

\section{Spin-One Heisenberg Chain  \label{pt}}

 It is  well known  that in the atomic limit, $U\gg t$,  the low-energy physics of the half-filled Hubbard  model  is described by the spin-half antiferromagnetic Heisenberg model. This result can be derived via second-order  perturbation theory.\cite{Auerbach}  In the same spirit, we now show that the low energy physics of two thirds filled Hubbard model on the triangular necklace model is described by an effective spin-one Heisenberg model  in the   molecular limit,  $t\ll t_c, U$.

\begin{figure}
 \includegraphics[width=0.99\columnwidth]{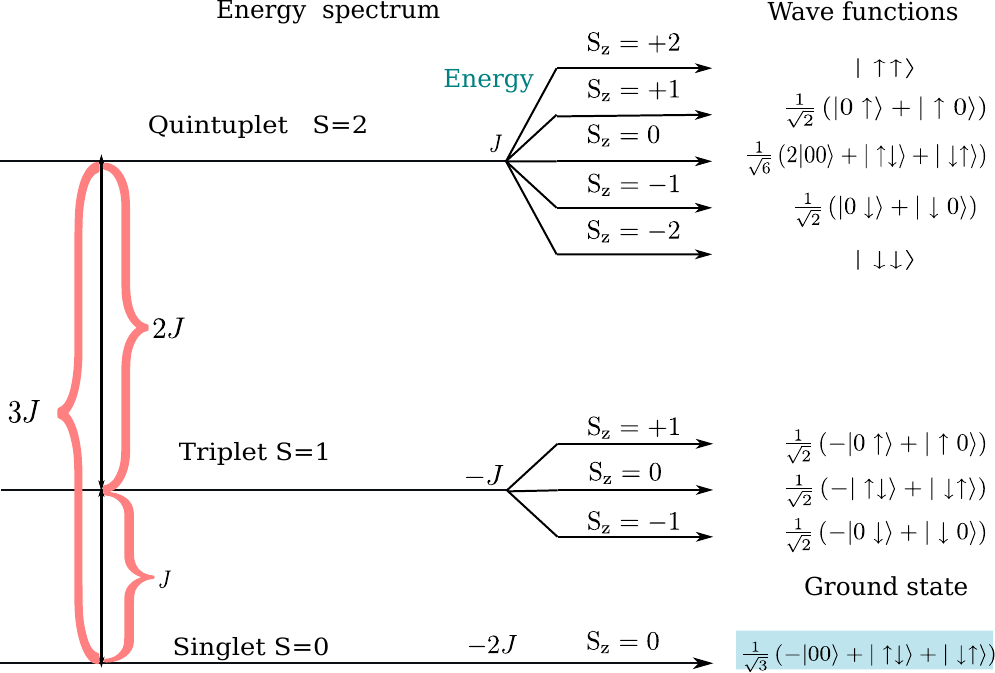}
\caption{Spectrum of  the two site Haldane model, $\hat{\cal H}=J\hat{\bf S}_1\cdot\hat{\bf S}_1$. Here, the local basis on each site is represented by $|0\rangle$, $|\ua \ra$ and $|\da \ra$ corresponding to $S_z$=0, $S_z$=1 and $S_z$=-1 respectively.}
\label{haldane2}
\end{figure}

  We take $\hat{\mathcal{H}}_\textrm{m}$ ({\it cf}. \eref{Ham:orb}) as  the zeroth order Hamiltonian and include $\hat{\mathcal{H}_t}$ ({\it cf}. \eref{perturbation}) perturbatively. As we will only work to second order we may limit our calculation to a system composed of two molecules without loss of generality.
The ground state of $\hat{\mathcal{H}}_\textrm{m}$ on two molecules (i.e., a dimer) has bare energy    $E^{(0)}=-4t_c+2U$ and is nine-fold degenerate.    
As the total spin of the two molecules is a good quantum number of the full Hamiltonian, $\hat{\mathcal{H}}$, it is helpful to consider each spin sector independently. At various points in the derivation we will need to relate the results of the perturbation theory to the exact solution of the two-site spin-one Heisenberg model; for convenience, we summarize this in  \fref{haldane2}.

\subsection{Singlet sector}

The $S=0$ member of the ground state manifold of $\hat{\mathcal{H}}_\textrm{m}$ on the dimer consisting of the $i^\textrm{th}$ and $j^\textrm{th}$ monomers, where $j=i\pm1$, is
\begin{eqnarray}
 | \phi_{ij}^{0}\ra  = \frac{1}{\sqrt{3}}\left(|\phi^{\Uparrow}_i\ra\otimes|\phi^{\Downarrow}_j\ra+|\phi^{\Downarrow}_i\ra\otimes |\phi^{\Uparrow}_j\ra  -|\phi^{0}_i\ra\otimes |\phi^{0}_j\ra\right).\nonumber\\ \label{ground_state}
 \end{eqnarray}
Note that it has same form  as the singlet ground state of a two site spin-one antiferromagnetic Heisenberg model (cf. \fref{haldane2}).   To second order in $\hat{\mathcal{H}}_t$, the energy  of the singlet state is
\begin{eqnarray}
 E^{(2)}_{S=0}&=&E^{(0)}+ \sum_{m_{0}}\frac{\la \phi_{ij}^{0}|\hat{\mathcal{H}}_t|m_{0}\ra\la m_{0}|\hat{\mathcal{H}}_t|\phi_{ij}^{0}\ra}{E^{(0)}-E_{m_{0}}} \nonumber\\
  &=&-4t_c+2U-3\sum_{n=1}^4\frac{4t^2}{9 a_n\left(3t_c+\varepsilon_n\right)},\label{singlet_energy}
 \end{eqnarray}
 where  $|m_{0}\rangle$ is the set of all possible intermediate wavefunctions which are formed by the hopping of electrons from one monomer to another, $\varepsilon_1=U$,  and for $n>1$
\begin{eqnarray}
\varepsilon_n&=&\frac23\left[ U+\xi\cos\left(\frac{\phi+2\pi n}{3} \right)\right],  
 \end{eqnarray}
where
\begin{eqnarray}
\xi &=&  \sqrt{U^2 + 27 t_c^2 },   
 \end{eqnarray}
 and
\begin{eqnarray}
\phi &=& \pi+\arccos\left(\left(\frac{U}{\xi}\right)^3\right).
 \end{eqnarray}
$a_1=3$ and for $n>1$
\begin{eqnarray}
a_n&=&2|\alpha_n|^2+|\beta_n|^2+1,    
\end{eqnarray}
where
\begin{eqnarray}
\alpha_n &=&\frac{-12 t_c^2 U-9 t_c^2 \varepsilon_n+U^2 \varepsilon_n-2 U \varepsilon_n^2+\varepsilon_n^3}{\sqrt{2} (U-\varepsilon_n) (3 t_c+U-\varepsilon_n) \varepsilon_n}    
 \end{eqnarray}
and
\begin{eqnarray}
\beta_n&=&\frac{U-3 t_c -\varepsilon_n }{U+3 t_c  -\varepsilon_n}.  
 \end{eqnarray}

\begin{figure}
\begin{center}
 \includegraphics[width=\columnwidth]{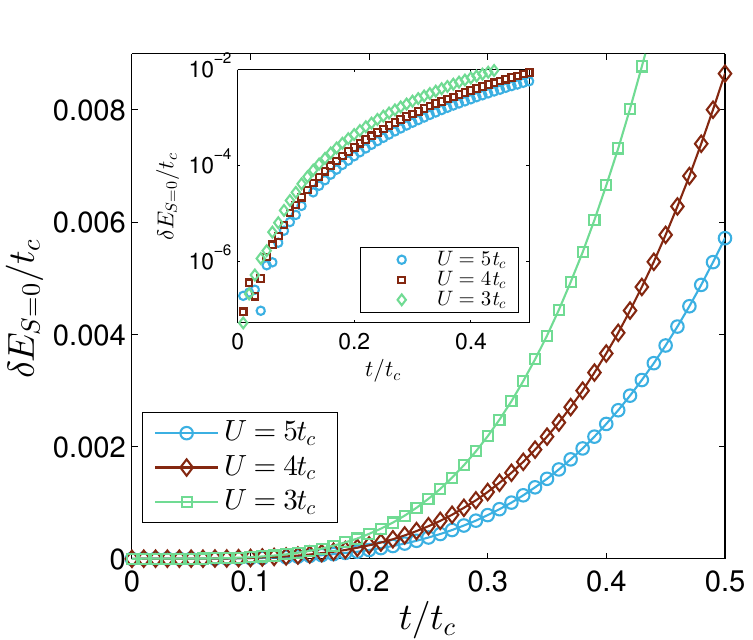}
\caption{The difference in energy $\delta E_{S=0}= |E_{S=0}^{(2)}-E^\textrm{DMRG}_{S=0}|$ of singlet wavefunctions   obtained from second order perturbation theory, $E_{S=0}^{(2)}$ and  DMRG, $E^\textrm{DMRG}_{S=0}$, for selected values of  $U$. Excellent agreement between the perturbation theory and the exact DMRG is found even for quite large $t$. 
The inset shows the same data on a semi-logarithmic scale.}
\label{s_0_perturbation}
\end{center}
\end{figure}

 In  \fref{s_0_perturbation},  we plot the  difference in the  energy,  $\delta E_{S=0}= |E_{S=0}^{(2)}-E^\textrm{DMRG}_{S=0}|$,  between the lowest energy  singlet wavefunctions obtained from  perturbation theory and the  DMRG ground state  for the  Hubbard model  on the two triangular dimer, which is a singlet. It can be  seen that, even for relatively small $U$, the error in the  energy obtained is of the $\mathcal{O}(10^{-3}$) for $t/t_c\lesssim0.25$ and $\mathcal{O}(10^{-2}$) for $t/t_c\lesssim0.5$. Thus, for the dimer, the perturbation theory gives remarkably good agreement with the DMRG results in the singlet sector.  %DMRG also shows that ground state  is a singlet, so the above  energy defined in  \eref{singlet_energy} is the groundstate energy  of the  Hubbard model on two triangular molecules at 2/3$^{rd}$ filling.  
 
\subsection{Triplet sector}

As spin is a good quantum number of the full Hamiltonian, we know that, while the perturbation may lift the degeneracy of the singlet, triplet and quintuplet, it will not split the triplet (or quintuplet). Therefore, it suffices to consider  only one of the unperturbed triplet wavefunctions of the dimer. A convenient choice (\textit{cf.} \fref{haldane2}) is
\begin{equation}
  |\phi_{ij}^{1}\ra  = \frac{1}{\sqrt{2}}\left( -|\phi^{\Uparrow}_{i}\ra\otimes|\phi^{\Downarrow}_{j}\ra+|\phi^{\Downarrow}_{i}\ra\otimes| \phi^{\Uparrow}_{j}\ra \right).
  \label{ground state:spin1}
 \end{equation}
 To second order in $\hat{\mathcal{H}}_t$, the energy of the triplet states is
 \begin{eqnarray}
E_{S=1}^{(2)}  &=&E^{(0)}-\frac{4t^2}{81t_c}-2\sum_{n=1}^4\frac{4t^2}{9 a_n\left(3t_c+\varepsilon_n\right)}\nonumber\\
  \label{triplet_energy}
 \end{eqnarray}

\begin{figure}
\begin{center}
 \includegraphics [width=\columnwidth]{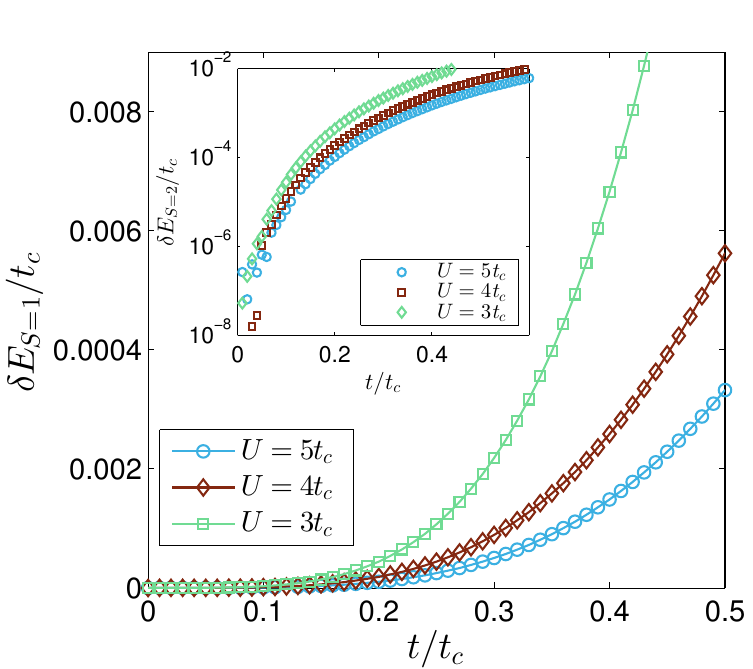}
\caption{The difference in the energy of the triplet wavefunction $\delta E_{S=1}$ obtained using perturbation theory and numerically using DMRG   for $U= 3tc,\;4t_c\;$ and  $5t_c$  as a function of perturbation $t/t_c$. Excellent agreement between the perturbation theory and the exact DMRG is found even for quite large $t$.
The inset shows the same data on a semi-logarithmic scale.}
\label{s_1_perturbation}
\end{center}
\end{figure}

 \fref{s_1_perturbation} shows the  of the energy difference,  $\delta E_{S=1}= |E_{S=1}^{(2)}-E^\textrm{DMRG}_{S=1}|$,  between the second order perturbation theory and the lowest energy triplet state found in the DMRG solution for the Hubbard model on a triangular necklace dimer. It can be seen that, as for the singlet case, the analytical calculation agrees well with the numerical results. 
 
%%%%%%%%%%%%%%%%%%%%%%%%%%5figure

\subsection{Quintuplet sector}

Finally, we consider the effect of perturbation on  one of the $S=2$ sector. It is convenient to consider the unperturbed state
\begin{equation}
 | \phi_{ij}^{2} \ra =  |\phi^{\Uparrow}_{i}\ra\otimes|\phi^{\Uparrow}_{j}\ra \label{ground state:spin2}
 \end{equation}
To second order   in $\hat{\mathcal{H}}_t$, the energy  of the quintuplet states is
\begin{equation}
E_{S=2}^{(2)}=E^{(0)}-\frac{4t^2}{27t_c}  \label{energys2}.
\end{equation}
We will see below that this simple form for the quintuplet energy is a consequence of the Pauli blockade. 

\begin{figure}
\begin{center}
 \includegraphics [width=\columnwidth]{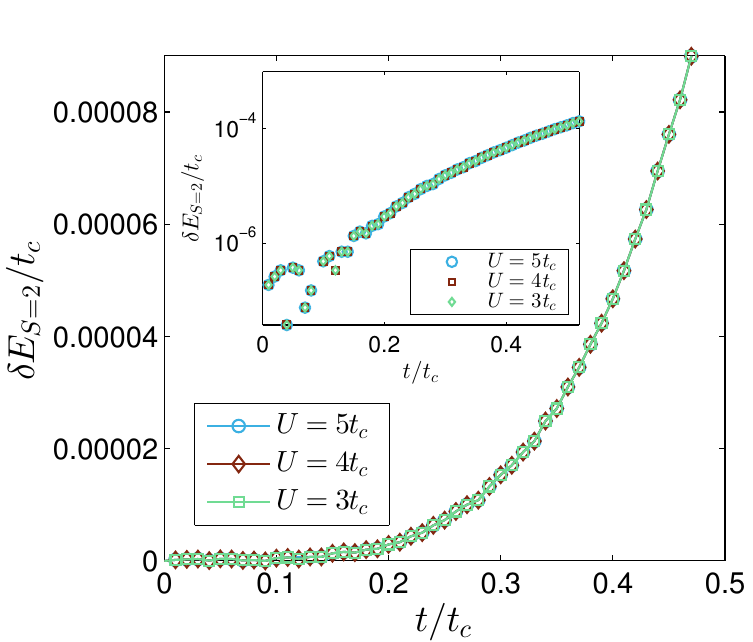}
\caption{The difference in the energy of the quintuplet  wavefunction $\delta E_{S=2}^2$ obtained using perturbation theory and from DMRG   for selected values of $U$. The error in the perturbation theory is two orders of magnitude smaller than for the singlet (Fig. \ref{s_0_perturbation}) or triplet (Fig. \ref{s_0_perturbation}) sectors. Furthermore, the error is independent of $U$. Both of these effects are to due to the Pauli blockade. 
The inset shows the same data on a semi-logarithmic scale.}
\label{s_2_perturbation}
\end{center}
\end{figure}

 \fref{s_2_perturbation}  shows difference in the energies, $\delta E_{S=2}= |E_{S=2}^{(2)}-E^\textrm{DMRG}_{S=2}|$, of the lowest quintuplet solutions found from second order perturbation theory and from DMRG. As for $S=0$ and $S=1$, the error in the perturbation theory is small: indeed for the quintuplet the errors are two orders of magnitude smaller than those for the singlet or triplet sectors. Yet the most striking feature of the plot is that the  the error in the energy  is independent of $U$. This is  a consequence of Pauli blockade and is simple to understand in the molecular orbital basis. In the unperturbed quintuplet state described by Eq. (\ref{ground state:spin2}) each molecule contains three spin-up electrons (one in each MO) and one spin-down (in the $A_+$ orbital). Therefore, to second order, the only possible corrections involve the spin-down electron virtually hopping into the $E_+$ orbital (recall that the local parity symmetry forbids hopping into the $E_-$ orbital). As these fluctuations do not change the total number of doubly occupied sites (indeed no processes can change the number of doubly occupied sites as we have six spin-up and two spin-down electrons in six orbitals and there are no spin flip terms in the perturbing Hamiltonian, $\hat{\mathcal{H}}_t$) $U$ cannot enter into the correction to the quintuplet energy, $E_{S=2}^{(2)}-E^{(0)}$. This results in the simple form of Eq. (\ref{energys2}). Indeed, the only higher order corrections on the dimer involve both down electrons taking part in such virtual process. This explains why both  $E_{S=2}^{(2)}-E^{(0)}$ and  $\delta E_{S=2}$, are independent of $U$ and why the perturbation theory is so accurate.

\subsection{Calculation of interaction strength, $J_s$ }

\begin{figure}
\begin{center}
 \includegraphics [width=\columnwidth]{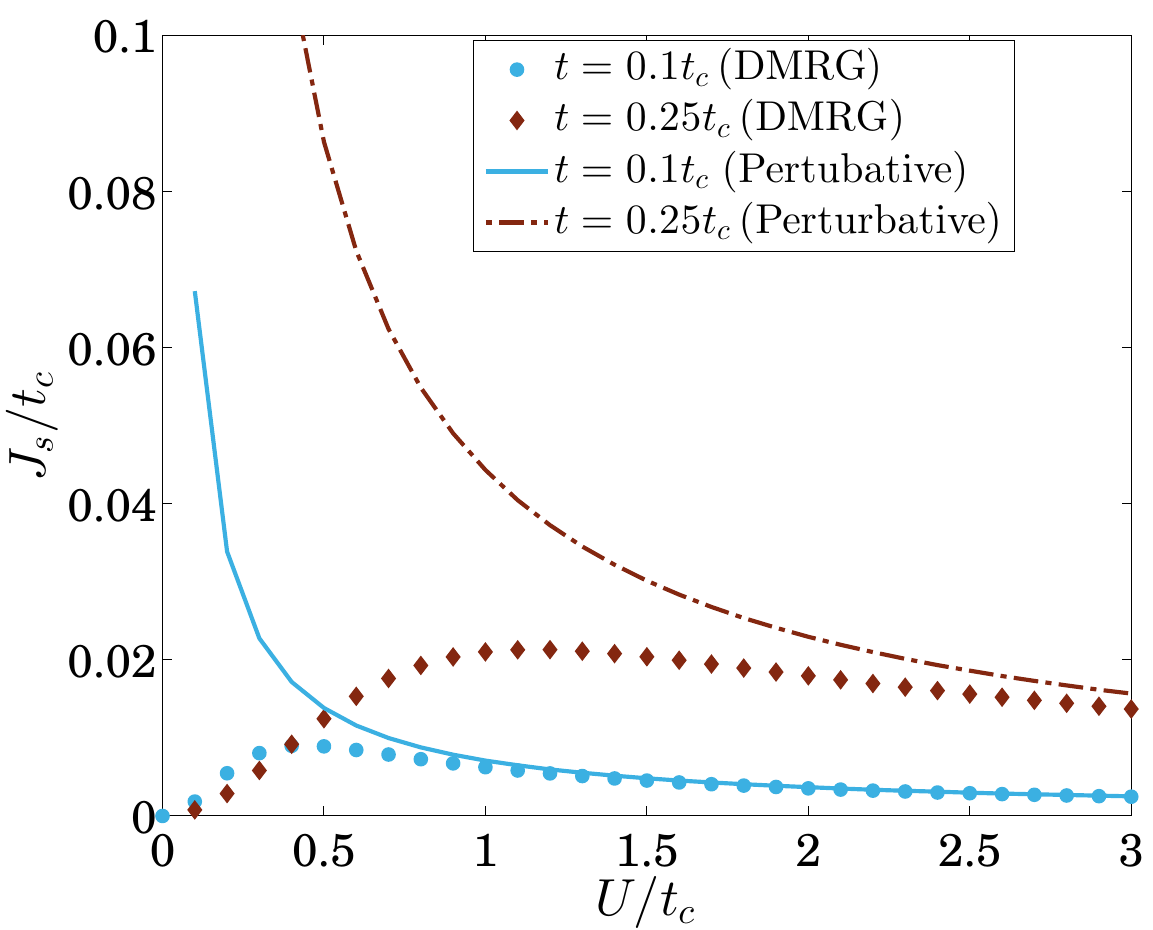}
\caption{Comparison of the interaction strength $J_s\equiv E_{S=1}-E_{S=0}=(E_{S=2}-E_{S=1})/2=(E_{S=2}-E_{S=0})/3$  as function of $U$ obtained from perturbation theory and from DMRG calculations for the Hubbard model on the six-site dimer.}
\label{interaction_strength}
\end{center}
\end{figure}

With the above results in hand, we can now make explicit connection to the Heisenberg model. 
The singlet-triplet energy gap is
\begin{eqnarray}
J_s \equiv E^{2}_{S=1}-E^{2}_{S=0}
\label{JS10}
%&=&-\frac{4t^2}{81t_c}+\frac{4t^2}{27 (3 t_c + U)}+\frac{4t^2}{ 9 a_1\left(3t_c+\varepsilon_1\right)}\nonumber%\\&&
%+\frac{4t^2}{ 9 a_2\left(3t_c+\varepsilon_2\right)}+\frac{4t^2}{ 9 a_3\left(3t_c+\varepsilon_3\right)}\\
=\sum_{n=0}^4\frac{4t^2}{9 a_n\left(3t_c+\varepsilon_n\right)}
\end{eqnarray}
where we have defined $\varepsilon_0=0$ and $a_0=-3$ for notational convenience.
 Similarly, we can calculate energy difference between  the singlet and quadruplet  and that of quintuplet and triplet. We find that
\begin{eqnarray}
 E^{2}_{S=2}-E^{2}_{S=1}=2 J_s\label{JS21}\\
E^{2}_{S=2}-E^{2}_{S=0}=3J_s \label{JS20}
\end{eqnarray}
 The above spectrum precisely maps onto  that of spin-1 Heisenberg dimer, summarised in  \fref{haldane2}. 
 A comparison of the interaction strength obtained from perturbation theory and DMRG is shown in \fref{interaction_strength}. The estimation of the interaction strength $J_s$ calculated using perturbation theory agrees well with the DMRG calculation in the limit  $U,\; t_c\gg t$ where one would expect the perturbation theory to hold.

\begin{figure}
 \begin{center}
  \includegraphics[width =0.95\columnwidth]{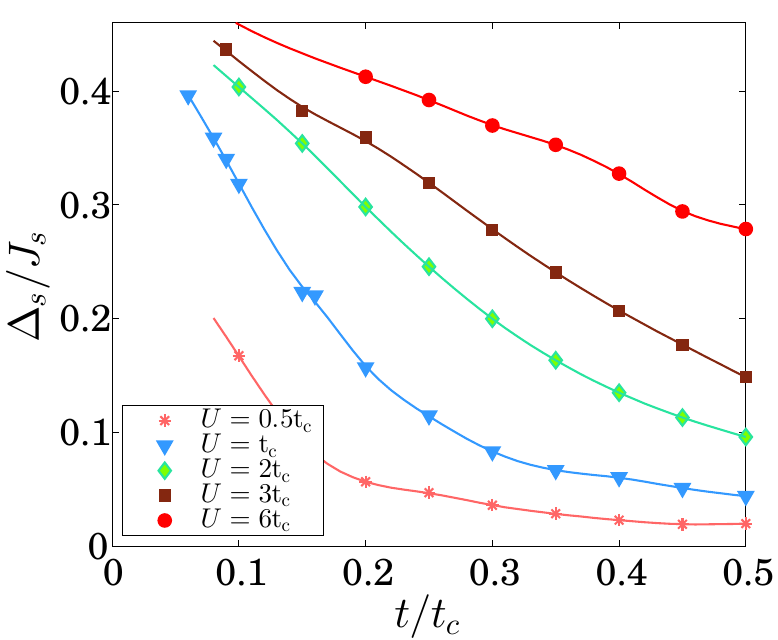}
\end{center}
\caption{
The spin gap,  $\Delta_s$, for $L=40$ molecules is consistent with the expected size of the Haldane gap expected from the magnitude of the superexchange interaction $J_s$. For comparison the Haldane spin gap for the spin-1 Heisenberg model is $0.41J_s$ \cite{White}.
Curves are guides to the eye.
\label{fig:spin-gap-scaled} 
}
\end{figure}

For DMRG calculations on large systems with open boundary conditions we find that the ground state is a singlet, but there is a triplet state at a vanishingly small energy above the ground state. These two states are separated from all other states by a much larger spin gap. This is consistent with the $D_2\cong Z_2\times Z_2$ degeneracy expected for the Haldane phase.\cite{Kennedy2,Kennedy1} This can be understood as arising from the the emergent spin-1/2 edge states, which have an interaction that becomes exponentially small as the system is taken into the thermodynamic limit.\cite{Ken}
In Fig. \ref{fig:spin-gap-scaled} we plot the spin gap, $\Delta_s=E_2(4L)-E_0(4L)$ where $E_S(N_e)$ is the energy of the lowest energy eigenstate for $N_e$ electrons in the spin-$S$ subspace, calculated from DMRG for $L=40$ molecules (120 sites), scaled by $J_s$ for a range of parameters. We find, as expected, that in the strong coupling molecular limit  the value of $\Delta_s/J_s$ is comparable to that found for the spin-1 Heisenberg model.\cite{White} Indeed the agreement is remarkably good given the additional numerical difficulties of dealing with a fermionic system such as the Hubbard model.

Thus  we   conclude that the spin-one antiferromagnetic Heisenberg model with interaction strength $J_s$ provides an effective low-energy theory of the two-thirds filled Hubbard model on the  triangular necklace  lattice in the limit $U,\; t_c\gg t$. It follows that, in this limit, the Hubbard model will be in the Haldane phase.\cite{Haldane1,Haldane2} However, this model does not give any insight into what happens outside of this limit, when one expects charge fluctuations to become important. Furthermore, this limit gives no insight into why the two-thrids filled Hubbard model on the triangular necklace model is insulating at finite $U$.

\section{Ferromagnetic Hubbard-Kondo lattice model  \label{kondo}}

% In the previous section, we have seen that in limit $U, t_c\gg t;\; t \rightarrow 0$, the ground state of HTNL  at   2/3$^rd$ filling  low lying excitations  exactly maps to the spin-1 anti-ferromagnetic Heisenberg model.  It can been show that for larger values of $t$, by projecting out $A_{+}$ orbitals, we can  map the HTNL to a correlated ferromagnetic coupled Kondo-lattice model(CFKLM) or the ferrom magnetic Kondo Hubabrd model at half-filling.  In particular, electrons in $E_{-}$  orbital act like localised electrons and $E_{+}$ orbital  behaves as the conduction electrons.

%In order to arrive at the effective Hamiltonian for the HTNL,  we take the conventional route of  projecting out some degrees of freedom in the Hamiltonian  based on certain symmetry arguments and signatures we see in our numerical calculation. 

In this section we show that, if the $A_+$ orbitals are projected out of the low-energy model we are left with a Hubbard-Kondo lattice model with the itinerant $E_+$ electrons ferromagnetically coupled to localised spins in the $E_-$ orbitals.

\subsection{Projection of the Hamiltonian onto the   $\hat n_{iE_{-} }=1$ subspace}

We have seen above that in the  molecular limit with $U=\infty$, there is exactly one electron in every $E_-$ orbital. Furthermore, we have seen that the local parity symmetry on every  molecule is a conserved quantity. Therefore, the occupation of the $E_-$ orbital is conserved modulo two. This means that, if we start from the strong coupling molecular limit and gradually reduce $U$ and increase $t$ one should expect the $E_-$ orbitals to remain singularly occupied unless or until there is a phase transition, where the occupation number of the $E_-$ orbitals may change non-adiabatically. This can occur because the absence of adiabatic continuity at a phase transition allows energy levels with different parities to cross.

\begin{figure}
\begin{center}
 \includegraphics [width=\columnwidth]{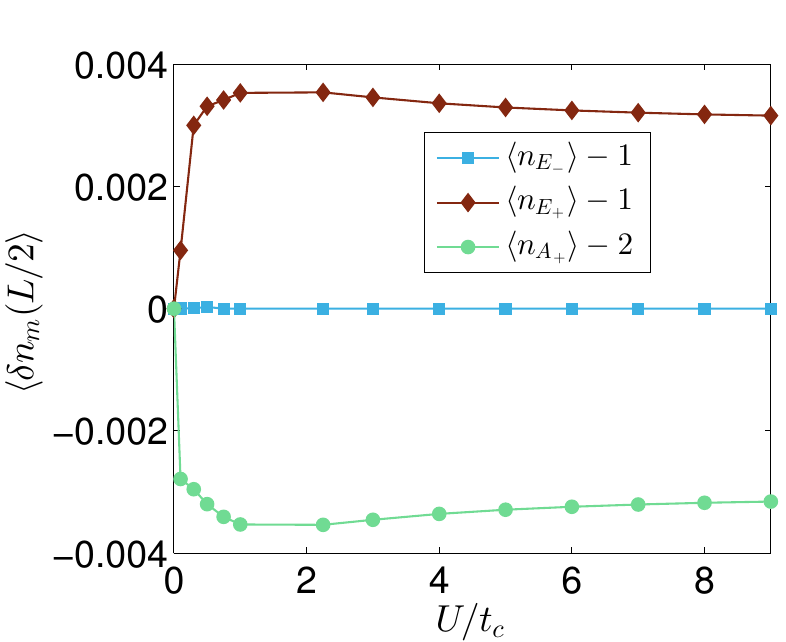}
\caption{The deviation in the filling of the molecular orbitals filling from that expected in the $U=\infty$ molecular limit.}
\label{orbital_filling}
\end{center}
\end{figure}

\begin{figure}
\begin{center}
 \includegraphics [width=\columnwidth]{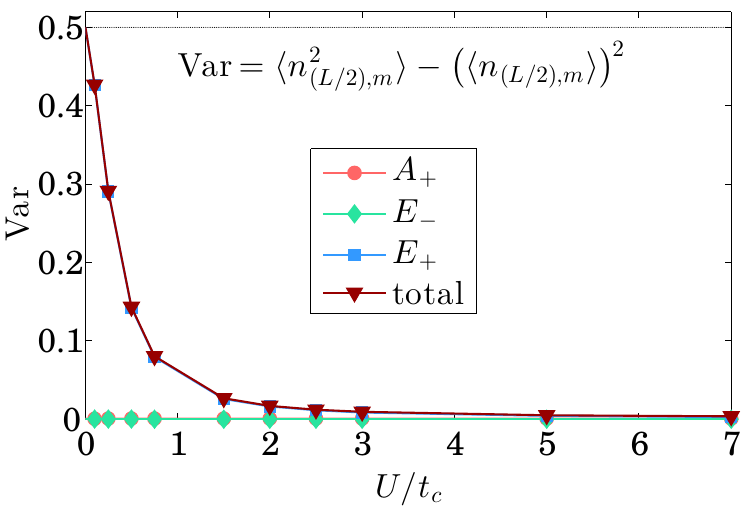}
\caption{ Charge fluctuations in different orbitals on a molecule of the chain  as a function of $U$ for $t=0.1t_c$ using DMRG. These calculations indicate that for all values of $U$ studied there are no charge fluctuations in the $E_-$ orbital, as expected from symmetry when $n_{E_-}=1$. The charge fluctuations in $A_{+}$ is also negligible.}
\label{variance}
\end{center}
\end{figure}

In \fref{orbital_filling}, we plot the  occupations  of the molecular orbitals for the Hubbard model on 120 sites (40 molecules) obtained from DMRG calculations. In Fig. \ref{variance} we plot the variance in these occupations.  We can see that for all $U$, $n_{A_{+}}\simeq 2$, $n_{E_{-}}=1$ and $n_{E_{+}} \simeq 1$. A striking feature of the above plot is that the occupancy of the $E_{-}$ orbital is strictly one. Furthermore, there are no charge fluctuations in $E_{-}$ orbital (cf. \fref{variance}). We have seen above that there is exactly one electron in each $E_-$ orbital for $U\rightarrow\infty$ and that, provided there is not phase transition this remains the case for smaller $U$. Therefore the numerical finding that  $n_{E_{-}}=1$ is consistent with the previous finding\cite{Janani} that there is no phase transition, at least down to a very small $U$ where numerics become extremely challenging, as $U$ is reduced on the triangular necklace model.

% The electron occupation  in $n_{iE_{-}}=1$ and no  charge fluctuation  is a  consequence of conservation of the parity symmetry by the Hamiltonian. In \sref{tb}, we showed that for $t=0$ and $U=0$, molecular orbitals have  a definite parity about the local reflection axis.  This parity is a symmetry of the Hamiltonian, as the Hamiltonian remains unchanged when we exchange sites 2 and 3 and  so it will be a conserved quantity.  As $i^\text{th}$ $E_{-}$ orbital is the only molecular orbital that is odd with respect to the local reflection of the $i^\text{th}$ molecule, it immediately follows that the number of electrons in any $E_{-}$ orbital can only increase or decrease by two.

%In the strong coupling  limit $U \rightarrow \infty$, (See appendix \ref{Nagaoka})  we show that occupation in the  $E_{-}$ orbital is $n_{iE_{-}}=1$ for all $i$. Moving away $U\rightarrow \infty$, towards finite $U$, the orbital occupation in $E_{-}$, $n_{iE_{-}}$ can only change from 1, if there is a phase transition in the charge sector,  which breaks the   local reflection symmetry of the Hamiltonian. Further, there are no interactions in the Hamiltonian which changes the occupation in $E_{-}$ orbital.  Thus for all $U>0$,   $n_{iE_{-}}=1$ is conserved separately for every molecule.  

Therefore,  we conclude that, at least in a large region of (and probably throughout) the phase diagram $\hat n_{iE_-}=1$ on all molecules and one can  project the Hamiltonian onto the subspace with exactly one electron in $E_{-}$ orbital on every molecule without introducing an approximation. % provides there is no phase transition in the charge sector. 
The projection operator onto $\hat n_{iE_{-}}=1$ is
\begin{equation}
\hat P_1=\left[1-\hat n_{iE_{-\uparrow}}\hat n_{iE_{-\downarrow}}\right]\left[\hat n_{iE_{-\uparrow}}+\hat n_{iE_{-\downarrow}}\right].
\label{PE}
\end{equation}
Under this projection the Hamiltonian $\hat{\cal H}$ yields
\begin{eqnarray}
\tilde{\cal H}=\hat P_{1}^{\dag}\hat{\cal H}\hat P_{1}
=\hat P_1^\dagger\left(\tilde{\cal H}_{t}+\tilde{\cal H}_{1}+\tilde{\cal H}_{2}+\tilde{\cal H}_{3}\right)\hat P_1 \label{ham:proj},
\end{eqnarray}
where,
\begin{subequations}
\begin{eqnarray}
\tilde{\cal H}_{t}
&=&\sum_{\sigma}\sum_{n,m\ne E_-}\left(\hat c^{\dag}_{im\sigma}T_{mn}\hat c_{(i+1)n\sigma}+H.c.\right),\\
\tilde{\cal H}_{1}
&=&\sum_{m\sigma}\varepsilon_m \hat c^\dagger_{im\sigma} \hat c_{im\sigma}+\sum_{m\ne E_{-}} U_m \hat n_{im\ua} \hat n_{im \da},\\
\tilde{\cal H}_{2}
&=&\sum_{imn}J_{mn}\hat {\mathbf{S}}_{im}.\hat {\mathbf{S}}_{in}+\sum_{imn\sigma}V_{mn }\hat {n}_{im\sigma}\hat {n}_{in\sigma'}\nonumber\\
        &&+\sum_{i\sigma}\sum_{m,n\ne E_{-}}X_{mn}\left(\hat n_{im\sigma} \hat c^\dagger_{im\bar{\sigma} } \hat c_{in\bar{\sigma}}+H.c.\right)\nonumber\\
       &&+\sum_{i}\sum_{m,n\ne E_{-}}P_{mn} \hat c^\dagger_{i m \ua } \hat c^\dagger_{i m\da}\hat c_{i n\ua} \hat c_{i n\da},
\end{eqnarray} 
 and 
\begin{eqnarray}
 \tilde{\cal H}_{3}&=&\frac{U}{3 \sqrt{2}}\sum_{i\sigma}\left(\hat  c^\dagger_{i E_{-}\sigma}\hat c_{i E_{-}\overline\sigma}  \hat c^\dagger_{i E_{+}\overline\sigma} \hat c_{i A_{+}\sigma} + H.c.\right)\nonumber\\
&& -\frac{U}{3\sqrt{2}}\sum_{i\sigma}\left(\hat {n}_{i E_{-}\sigma} \hat c^\dagger_{i E_{+}\overline\sigma}\hat c_{i A_{+}\overline\sigma} +H.c.\right).
\end{eqnarray} 
\end{subequations}

\subsection{Projection of the Hamiltonian to   $\hat n_{iA_{+} }=2$} 

In the  molecular limit, \textit{i.e.}, as  $t\rightarrow 0$ ({\it cf}. sec \ref{MO}) we  found that the $A_+$ orbitals are doubly occupied. This is also the case in the non-interacting ($U=0$) and $U=\infty$ solutions. Away from these limits we expect  this to be an approximation as this is not protected by symmetry. The DMRG calculations in  Figs. \ref{orbital_filling} and \ref{variance} coincide with this expectation, but show that, for all $U$ studied, the charge transferred from the $A_+$ orbitals to the $E_+$ orbitals is negligibly  small. Furthermore, because these orbitals are nearly filled, rather than, say, nearly half-filled as is the case for the $E_+$ orbital, one does not  see large charge fluctuations in the $A_+$ orbital  and therefore one does not expect electronic correlations in the $A_+$ orbitals to play an important role in determining the physics of the Hubbard model. This is borne out by the DMRG calculations (see also \fref{variance}). Therefore we now further project onto the $\hat n_{iA_{+} }=2$ subspace  via a  {`anti-Gutzwiller'}  projection $\hat  P_{\overline G}= \hat n_{A_{+}\ua} \hat n_{A_{+}\da}$. On the basis of the analytical results described above we expect this approximation to work best for small $t$ and in both the limits $U\rightarrow0$ and $U\rightarrow\infty$. It is less clear how good this approximation remains for intermediate $U$ and large $t$ where the charge fluctuations in the $A_+$ orbital in our DMRG calculations are largest. 

Projecting the Hubbard model onto both $n_{E_{-}}=1$ and  $n_{A_{+}}=2$ one finds that
\begin{eqnarray}
\hat{H}_\textrm{eff}&\equiv&\hat{P}_{\overline G}^{\dag}\hat{P}_{1}^{\dag}\hat{H}\hat{P}_{1}\hat{P}_{\overline G} \label{ham:proj:A}\\
%\end{eqnarray}
%Now, replacing  $n_{E_{-}}=1$ and  $n_{A_{+}}=2$, we arrive at the effective Hamiltonian for the HTNL at 2/3$^{rd}$ filling  as below
%\begin{eqnarray}
% H_{eff}
&=&N\varepsilon_{E_+}^* - t^*\sum_{i\sigma}\left(\C{iE_{+}\sigma}\A{i+1E_{+}\sigma}+H.c.\right)\notag\\
&&+U^* \sum_{i}\hat n_{iE_{+} \ua}\hat n_{iE_{+} \da}\nonumber%\\&&
-J^*\sum_{i}\hat {\mathbf{S}}_{iE_{+}}\cdot\hat {\mathbf{S}}_{iE_{-}}%\nonumber\\&&
%+\epsilon_1 \sum_{i \sigma} n_{i E_{+} \sigma}
%+N\epsilon_0\label{ham:kondo},%\nonumber\\
\end{eqnarray}
where %$\epsilon_1=2V_{A_{+} E_{+}}+ V_{E_{+} E_{-}}=5U/12$, $\epsilon_0=2\varepsilon_{A_{+}}+ \varepsilon_{E_{-}}+\varepsilon_{E_{+}} + U_{A_{+}}=????$,
   $\varepsilon_{E_+}^*=4V_{A_{+} E_{+}}+ V_{E_{+} E_{-}}+2\varepsilon_{A_{+}}+ 2\varepsilon_{E_+} +U_{A_{+}}=17U/24-2t_c$,  $t^*=-T_{E_+E_+}=2t/3$, $U^*=U_{E_+}=U/2$, $J^*=-2J_{E_{+} E_{-}}=U/3$,
and $N$ is the  total number of  molecules.
Up to the trivial term proportional to $\varepsilon_{E_+}^*$, this is simply the Kondo lattice model with a ferromagnetic exchange interaction between the localized $E_-$ spins and the itinerant $E_+$ electrons with a on site repulsive  Hubbard interaction between the $E_+$ electrons, i.e.,  the ferromagnetic Hubbard-Kondo lattice model.

%Thus,  we can see from \eref{ham:kondo}, the Hamiltonian is effectively a strongly correlated  ferromagnetic coupled Kondo-lattice model at half filling with $E_{-}$ orbital being the localised electrons and $E_{+}$ electrons the conduction electron.

The ferromagnetic Hubbard-Kondo lattice model with $S=1/2$ impurities has not been extensively studied at half-filling in one spatial dimension.  A numerical study using quantum Monte Carlo (QMC), DMRG and exact diagonalization of this model for large $U^*$  and  various dopings,\cite{Dagotto} found that model has a complicated phase diagram, with a ferromagnetic phase away from half filling and incommensurate spin order  near  half filling.   However,  at half filling, the nature of the ground state is not clear, as the half filling density is not accessible to QMC  due to sign problem. %In fact, there are no further investigation on this model exactly at half filling, but we can infer the ground state of this model from related  models. 
 
The $U^*=0$ version of this model, i.e., the ferromagnetic Kondo lattice model has been studied in more detail for various doping. \cite{Garcia,Tsunetsugu,Dagotto} But, again, the half-filled case has received scant attention. We are only aware of two very brief reports, \cite{Garcia,Tsunetsugu} which claim that this model is insulating, with antiferromagnetic correlations with a spin gap and has a ground  state which belongs to the Haldane phase. Clearly this is correct for $J^*\rightarrow\infty$ in our model.  Further, an investigation of a variant of the ferromagnetic Hubbard-Kondo lattice model with additional interactions,\cite{Malvezzi} found that onsite Coulomb interactions does not change the phase of the model qualitatively.   Yanagisawa and Shimoi \cite{Yanagisawa}  proved that for a bipartite lattice with $U^*>J^*/4$ the ground state of  the  ferromagnetic Hubbard-Kondo lattice model  is a singlet. This is consistent with our results, which correspond to the relevant parameter regime. (Note that although the triangular necklace model is frustrated the ferromagnetic Hubbard-Kondo model defined by Eq. (\ref{ham:proj:A}) lives on a bipartite lattice.)

This suggests that  $J^*$, rather than $U^*$, is the  physically important interaction in the ferromagnetic Hubbard-Kondo lattice model. This provides a simple physical picture for the insulating state in both the ferromagnetic Kondo lattice model and the full Hubbard model on the triangular necklace lattice. Specifically, the formation of Kondo triplets confines itinerant $E_+$ electrons. Therefore the insulating state is best understood as a (ferromagnetic) Kondo insulator, with the formation of triplets being responsible for the localisation of the itinerant electrons, rather than a Mott insulator.  This is consistent with our finding that the large $U$ limit of the model is in the Haldane phase, rather than the Luttinger liquid phase that would be expected for the spin degrees of freedom if the $E_+$ electrons formed a Mott insulator.

 \section{Conclusion}
 
 We have shown that,  in the molecular limit, the low-energy physics of the two-thirds filled Hubbard model on the triangular necklace lattice  is described by the spin-one Heisenberg chain. Away from the molecular limit the low-energy excitations of this model is well approximated by the ferromagnetic Hubbard-Kondo lattice model. This gives a natural explanation of the unexpected insulating state recently discovered  for the two-thirds filled  Hubbard model on the triangular necklace lattice, {\it viz} that it is a (ferromagnetic) Kondo insulator. The Haldane phase found for the two-thirds filled Hubbard model on the triangular necklace lattice is consistent with previous arguments that the  ferromagnetic Hubbard-Kondo lattice model has a Haldane ground state. 
 
 We have also shown that Hund's rules for a three site `molecule' share the same physical origin as Nagaoka's theorem.

 %We have investigated the Hubbard model on triangular necklace lattice at $2/3^{rd}$ filling using second order perturbation theory and DMRG, and show that low lying spin excitations is exactly same as that of a two site Heisenberg model in large $U$, small $t$ limit and ground state is in Haldane phase. We have also shown tha HTNL $2/3^{rd}$ can be mapped on to the ferromagnetically coupled Kondo lattice Hubbard model with spin half impurities at half filling whose ground state is also in Haldane phase. These calculation also show how HTNL is an insulator at $2/3^{rd}$ filling (More to be written based on the write up by Ben).

\section*{Acknowledgments}

This work was supported by the Australian Research Council (grants DP0878523, DP1093224,  LE120100181, and FT130100161) and MINECO (MAT2012-37263-C02-01). This research was undertaken with the assistance of resources provided at the NCI National Facility through the National Computational Merit Allocation Scheme supported by the Australian Government and was supported under Australian Research Council's LIEF funding scheme (project number LE120100181).


\begin{thebibliography}{100}
\bibitem{foot-sign} Note that Tasaki\cite{Tasaki} takes the opposite sign convention to us, i.e., our negative $t$ corresponds to his positive $t$.
\end{thebibliography}
\end{document}